\documentclass[twocolumn,showpacs,preprintnumbers,amsmath,amssymb]{revtex4}

\usepackage{graphicx}

\begin{document}

\title{
Nonradial stability of marginal stable circular orbits 
in stationary axisymmetric spacetimes 
}
\author{Toshiaki Ono}
\author{Tomohito Suzuki}
\author{Hideki Asada} 
\affiliation{
Faculty of Science and Technology, Hirosaki University,
Aomori 036-8561, Japan} 

\date{\today}

\begin{abstract}
We study linear nonradial perturbations and stability of a marginal stable circular orbit (MSCO) such as the innermost stable circular orbit (ISCO) of a test particle in stationary axisymmetric spacetimes which possess a reflection symmetry with respect to the equatorial plane. A zenithal stability criterion is obtained in terms of the metric components, the specific energy and angular momentum of a test particle. The proposed approach is applied to Kerr solution and Majumdar-Papapetrou solution to Einstein equation. Moreover, we reexamine MSCOs for a modified metric of a rapidly spinning black hole that has been recently proposed by Johannsen and Psaltis [PRD, {\bf 83}, 124015 (2011)]. We show that, for the Johannsen and Psaltis's model, circular orbits that are stable against radial perturbations for some parameter region become unstable against zenithal perturbations. This suggests that the last circular orbit for this model may be larger than the ISCO. 
\end{abstract}

\maketitle

\section{Introduction}
GW150914 event that has been eventually discovered by the two aLIGO detectors 
fits well with a binary black hole inspiraling and merger model 
\cite{GW150914PRL,GW150914ApJL}. 
It is likely that there exists the innermost stable circular orbit (ISCO) 
for the binary black hole. 
In gravitational wave astronomy, 
ISCOs are thought to be the last circular orbit 
corresponding to the transition point from the 
inspiraling phase to the merging one. 
In the post-Newtonian approximation, the dynamical stability of circular 
orbits of a two-body system has been studied. 
See Ref. \cite{Blanchet} for a comprehensive review 
on the post-Newtonian approximation: 
The 1PN term gives exactly the ISCO radius for the Schwarzschild case, 
but for any mass ratio. 
The finite mass corrections on the ISCO radius appear at the 2PN order. 
At the 3PN order, all the circular orbits are stable 
when the mass ratio is larger than some critical value $\nu_C \sim 0.183$ 
(See e.g. Section 8.2 in \cite{Blanchet}). 
However, there is an ISCO at the 3PN order, when the mass ratio 
is smaller than the critical value. 
Moreover, at the 4PN (and higher PN) orders, finite mass corrections
on the ISCO radius may appear \cite{Blanchet-priv}. 
In black hole perturbation theory, 
Isoyama et al. have discussed the ISCO of a small particle 
around a Kerr black hole and have 
computed the linear-order correction in the mass ratio 
\cite{Isoyama}.

In high energy astrophysics, furthermore, 
ISCOs are related to 
the inner edge of 
an accretion disk around a black hole 
\cite{Abramowicz}. 
Therefore, the ISCOs 
would help us to test the nonlinear spacetime geometry 
that is beyond the solar-system experiments. 
It is important to study the nature of the ISCOs 
around black holes 
with electric charges and/or scalar fields 
\cite{no-hair} 
and black holes in modified gravity theories 
\cite{modified}.

Ono et al. have recently reexamined, 
in terms of Sturm's theorem \cite{Algebra},  
marginal stable circular orbits (MSCOs) 
of a test particle in spherically symmetric 
and static spacetimes \cite{Ono}. 
The wording ``marginal stable circular orbits'' 
is used, because the wording ``marginally stable'' has 
a different meaning in the theory of dynamical systems. 
It is interesting to extend such previous works 
\cite{no-hair,modified,Ono} to stationary and axisymmetric cases, 
because astrophysical black holes are likely to spin 
\cite{GW150914PRL,GW150914ApJL}. 
Indeed, Stute and Camenzind have found 
an equation for the MSCO radius \cite{Stute}, 
where the particle is assumed to move on the equatorial plane.  
However, their analysis is limited within the radial perturbation. 

The main purpose of this paper is to extend such a MSCO study 
to stationary and axisymmetric spacetimes, 
and furthermore, 
by taking account of not only the radial perturbation but also 
the nonradial perturbation. 
In particular, a zenithal stability 
criterion for a test particle 
in a MSCO is derived in terms of 
the metric components, the specific energy and angular momentum 
of a test particle. 
We shall apply the proposed approach to three models: 
Kerr solution, 
Majumdar-Papapetrou solution to the Einstein equation 
\cite{MP1,MP2}, 
and Johannsen and Psaltis's model of 
a rapidly spinning black hole in modified gravity \cite{JP}. 

Throughout this paper, we use the unit of $G=c=1$. 

\section{Marginal stable circular orbit}
We consider a stationary axisymmetric spacetime. 
The line element for this spacetime is \cite{Lewis,LR,Papapetrou}
\begin{align}
ds^2=&g_{\mu\nu}dx^{\mu}dx^{\nu} \notag\\
=&-A(y^p,y^q)dt^2-2H(y^p,y^q)dtd\phi \notag\\
&+F(y^p,y^q)(\gamma_{pq}dy^pdy^q)+D(y^p,y^q)d\phi^2 , 
\label{ds-0}
\end{align}
where $\mu, \nu$ run from $0$ to $3$, 
$p, q$ take $1$ and $2$, 
$t$ and $\phi$ coordinates are associated with the Killing vectors,  
and $\gamma_{pq}$ is a two-dimensional symmetric tensor. 
It is more convenient to 
reexpress this metric into 
a form in which $\gamma_{pq}$ is diagonalized. 
The present paper prefers the polar coordinates 
rather than the cylindrical ones, 
because 
we consider 
the Kerr metric and its modified metric 
by Johannsen and Psaltis \cite{JP} 
in the polar coordinates. 
Then, Eq. (\ref{ds-0}) becomes \cite{Metric}
\begin{align}
ds^2=&-A(r,\theta)dt^2-2H(r,\theta)dtd\phi \notag\\
&+B(r,\theta)dr^2+C(r,\theta)d\theta^2+D(r,\theta)d\phi^2 . 
\label{ds}
\end{align}

The present paper focuses on 
a test particle 
outside the horizon. 
Then, $\det{(g_{\mu\nu})} < 0$. 
This leads to $B(r,\theta) \neq 0$, $C(r,\theta) \neq 0$ 
and $A(r,\theta)D(r,\theta)+H^2(r,\theta) \neq 0$. 
We assume also a local reflection symmetry with respect to 
the equatorial plane as \cite{Reflection}
\begin{equation}
\left. \frac{\partial g_{\mu\nu}}{\partial\theta}\right|_{\theta=\pi/2} = 0 ,  
\end{equation}
where we choose $\theta = \pi/2$ as the equatorial plane. 

We denote the contravariant metric tensor as 
\begin{align}
{g}^{\mu\nu}&=\left(
    \begin{array}{cccc}
      -\frac{D}{AD+H^2} & 0 & 0 & -\frac{H}{AD+H^2}\\
      0 & \frac{1}{B} & 0 & 0 \\
      0 & 0 & \frac{1}{C} & 0 \\
      -\frac{H}{AD+H^2} & 0 & 0 & \frac{A}{AD+H^2}
    \end{array}
    \right) \notag \\
&\equiv \left(
    \begin{array}{cccc}
      -\tilde{A}&0&0&-\tilde{H}\\
      0&\tilde{B}&0&0\\
      0&0&\tilde{C}&0\\
      -\tilde{H}&0&0&\tilde{D}
    \end{array}
    \right)　. 
\label{inverse}
\end{align}

Let us define the Lagrangian of the test particle as 
\begin{align}
{\cal L} = -A \dot{t}^2 -2H \dot{t}\dot{\phi} +B \dot{r}^2 
+C \dot{\theta}^2 +D \dot{\phi}^2 , 
\label{Lag}
\end{align}
where the dot denotes the derivative with respect to the proper time. 
There are two constants of motion for the test body, 
because the spacetime is stationary and axisymmetric.  
\begin{align}
\varepsilon 
&=-A \dot{t}-H \dot{\phi} ,
\label{e}\\
\ell 
&=-H\dot{t}+D\dot{\phi} , 
\label{l}
\end{align}
where $\varepsilon$ and $\ell$ are corresponding to the specific energy 
and the specific angular momentum, respectively. 

Let us examine linear perturbations around the equatorial 
marginal circular orbit (denoted as $r=r_c$), 
in which the perturbed position is denoted as $r=r_c +\delta r$ and 
$\theta = \pi/2 + \delta\theta$. 
We consider the geodesic equation at the perturbed position 
and 
examine the linear parts 
in $\delta r$ and $\delta\theta$. 
Straightforward calculations show that 
$\delta r$-parts and $\delta \theta$-ones 
are decoupled with each other in the geodesic equation. 

First, we consider the $r$-component of the geodesic equation 
in the Taylor series approximation such as 
$A(r, \theta) = A(r_c, \pi/2) 
+ \delta r \times \partial A(r, \pi/2)/\partial r|_{r=r_C} + \cdots$, 
where we use the local reflection symmetry of the metric with respect to 
the equatorial plane, and $\dot{t}$ and $\dot{\phi}$ are rewritten 
in terms of $\varepsilon$ and $\ell$ by using Eqs. (\ref{e}) and (\ref{l}). 
Hence, we obtain 
\begin{align}
\left.\frac{\partial \tilde{A}}{\partial r}\right|_c\varepsilon^2+2\left.\frac{\partial \tilde{H}}{\partial r}\right|_c\varepsilon l-\left.\frac{\partial \tilde{D}}{\partial r}\right|_cl^2=0 , 
\label{r-1}
\end{align}
at the zeroth order (namely, $r=r_C$ and $\theta=\pi/2$) 
and 
\begin{align}
\left.\frac{\partial^2 \tilde{A}}{\partial r^2}\right|_c\varepsilon^2+2\left.\frac{\partial^2 \tilde{H}}{\partial r^2}\right|_c\varepsilon l-\left.\frac{\partial^2 \tilde{D}}{\partial r^2}\right|_c l^2=0 , 
\label{r-2}
\end{align}
at the linear order, 
where $|_C$ denotes the position at $r=r_C$ and $\theta=\pi/2$. 

Next, the $\theta$-component of the geodesic equation 
gives us a stable condition to the zenithal perturbation as  
\begin{align}
\left.\frac{\partial^2 \tilde{A}}{\partial \theta^2}\right|_c\varepsilon^2+2\left.\frac{\partial^2 \tilde{H}}{\partial \theta^2}\right|_c\varepsilon l - \left.\frac{\partial^2 \tilde{D}}{\partial \theta^2}\right|_c l^2 < 0 , 
\label{theta}
\end{align}
which 
plays a crucial role 
in this paper. 
The inequality sign changes if the orbit is unstable 
against the zenithal perturbation. 
Henceforth, we omit the notation $|_C$ for brevity. 
See e.g. page 362 in Chandrasekhar (1998) for a description 
of the $\theta$-motion 
in a well-known black hole solution such as Kerr solution 
\cite{Chandrasekhar}. 
Note that 
a zenithal stability criterion by Eq. (\ref{theta}) 
is new.

By combining Eqs. (\ref{r-1}) and (\ref{r-2}) 
in the similar manner to the spherically symmetric case 
(See e.g. \cite{Ono,BJS,RZ}), 
we obtain
\begin{align}
&\left[ \frac{\partial\tilde{D}}{\partial r}\frac{\partial^2\tilde{A}}{\partial r^2}-\frac{\partial\tilde{A}}{\partial r}\frac{\partial^2\tilde{D}}{\partial r^2} \right]^2 
\notag\\
&-4\left[\frac{\partial\tilde{A}}{\partial r}\frac{\partial^2\tilde{H}}{\partial r^2}-\frac{\partial\tilde{H}}{\partial r}\frac{\partial ^2\tilde{A}}{\partial r^2} \right] \left[ \frac{\partial\tilde{D}}{\partial r}\frac{\partial^2\tilde{H}}{\partial r^2}-\frac{\partial\tilde{H}}{\partial r}\frac{\partial^2\tilde{D}}{\partial r^2}\right]=0 . 
\label{MSCO-eq}
\end{align}
After straightforward calculations, one can show that 
this agrees with the Stute and Camenzind's equation 
for the MSCO radius \cite{Stute}. 
Beheshti and Gasperin \cite{Beheshti} rediscovered in a mathematical way 
the Stute and Camenzind's equation, 
which is, to be precise, a necessary condition for the existence of MSCOs. 
Note that Eq. (\ref{MSCO-eq}) is shorter. 
See Eq. (40) in \cite{Stute}. 
This is only because we use the contravariant metric components. 
We call Eq. (\ref{MSCO-eq}) a MSCO equation for brevity. 
See Appendix for derivations of Eq. (\ref{MSCO-eq}).

Before closing this section, let us write down the expression 
for the squared specific energy and the squared specific angular momentum as 
\begin{align}
&\varepsilon^2 = 
\notag\\
&
\frac{\tilde{A}\left\{\frac{-\frac{\partial\tilde{H}}{\partial r} \pm \sqrt{(\frac{\partial\tilde{H}}{\partial r})^2+\frac{\partial\tilde{A}}{\partial r}\frac{\partial\tilde{D}}{\partial r}}}{\frac{\partial\tilde{A}}{\partial r}}\right\}^2}{\left[\frac{\tilde{A}}{\frac{\partial\tilde{A}}{\partial r}} \left\{-\frac{\partial\tilde{H}}{\partial r} \pm \sqrt{\left(\frac{\partial\tilde{H}}{\partial r}\right)^2+\frac{\partial\tilde{A}}{\partial r}\frac{\partial\tilde{D}}{\partial r}}\right\}+\tilde{H}\right]^2-\tilde{H}^2-\tilde{A}\tilde{D}} ,  
\label{e2}
\end{align}
\begin{align}
&\ell^2 = 
\notag\\
&
\frac{\tilde{A}}{\left[\frac{\tilde{A}}{\frac{\partial\tilde{A}}{\partial r}} \left\{-\frac{\partial\tilde{H}}{\partial r} \pm \sqrt{\left(\frac{\partial\tilde{H}}{\partial r}\right)^2+\frac{\partial\tilde{A}}{\partial r}\frac{\partial\tilde{D}}{\partial r}}\right\}+\tilde{H}\right]^2-\tilde{H}^2-\tilde{A}\tilde{D}} . 
\label{l2}
\end{align}
Note that Eqs. (\ref{e2}) and (\ref{l2}) hold 
for not only MSCOs but also general circular orbits.

\section{Applications to stationary axisymmetric models} 
\subsection{Kerr solution}
\begin{figure}
\includegraphics[width=8cm]{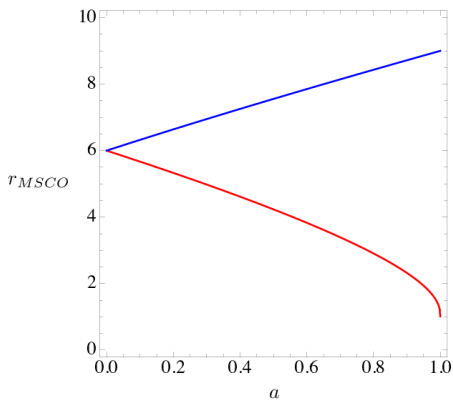}
\caption{ 
MSCOs around a Kerr black hole. 
The vertical axis denotes the ISCO radius in the unit of $M$. 
The horizontal axis denotes the spin parameter $a$. 
The upper curve (blue in color) corresponds to the retrograde case 
and the lower curve (red in color) corresponds to the prograde one. 
}
\label{fig-Kerr}
\end{figure}

Let us 
consider 
the Kerr metric 
\begin{align}
ds^2=&-\left(1-\frac{2Mr}{\Sigma}\right)dt^2-\frac{4aMr\sin^2\theta}{\Sigma}dtd\phi 
\nonumber \\
&+\frac{\Sigma}{\Delta}dr^2+\Sigma d\theta^2
\nonumber \\
&+\left(r^2+a^2+\frac{2a^2Mr\sin^2\theta}{\Sigma}\right)\sin^2\theta d\phi^2 , 
\label{Kerr}
\end{align}
where $\Sigma=r^2+a^2\cos^2\theta$ and $\Delta=r^2-2Mr+a^2$. 
Then, Eq. (\ref{MSCO-eq}) becomes  
\begin{align}
\overline{r}^4-12\overline{r}^3+(-6\overline{a}^2+36)\overline{r}^2-28\overline{a}^2\overline{r}+9\overline{a}^4=0 ,
\label{MSCO-eq-Kerr}
\end{align}
where $r$ and $a$ are normalized as $\bar{r} \equiv r/M$ 
and $\bar{a} \equiv a/M$. 
This quartic equation is solved to recover the known result. 
See Eq. (2.21) in \cite{BPT}. 
The zenithal stable condition by Eq. (\ref{theta}) is satisfied. 
Therefore, 
the same radius holds against the zenithal perturbation. 
See Figure \ref{fig-Kerr} for the ISCO location.

\subsection{Majumdar-Papapetrou solution}
Next, we consider the Majumdar-Papapetrou solution to Einstein equation 
\cite{MP1,MP2}. 
This solution expresses a superposition of 
extreme Reissner-Nordstr\"{o}m solutions. 
The spacetime metric is 
\begin{align}
ds^2&=-\Omega^{-2}dt^2+\Omega^2(dx^2+dy^2+dz^2) , 
\label{ds-MP}
\end{align}
where we define 
\begin{align}
\Omega=&1+\sum_{i} \frac{m_i}{r_i} ,
\notag\\
r_i=&\sqrt{(x-x_i)^2+(y-y_i)^2+(z-z_i)^2} . 
\end{align}

We focus on a case of two equal-mass black holes, 
where they are located at $(0, 0, L)$ and $(0, 0, -L)$ 
along the $z$ axis. 
By using the polar coordinates, 
Eq. (\ref{MSCO-eq}) 
for this equal-mass case becomes 
\begin{align}
&r^6+r^4 \left(6 L^2-6 M \sqrt{L^2+r^2}\right) 
\notag\\
&+ r^2 \left(9 L^4+10 L^2 M \sqrt{L^2+r^2}\right) 
\notag\\
&+ 16 L^4 M \sqrt{L^2+r^2}+4L^4(L^2+4M^2)=0 , 
\label{MSCO-eq-MP}
\end{align}
where $M$ is the black hole mass. 
By introducing $s \equiv (r^2+L^2)^{1/2}$, 
this long 
equation is reduced to 
\begin{equation}
s^6 - 6Ms^5 + 3 L^2 s^4 + 22 L^2 M s^3 + 16 L^4 M^2 = 0 . 
\label{MSCO-eq-MP2}
\end{equation}
Exact solutions for a sixth-degree equation are not known. 
Therefore, 
we have to solve numerically Eq. (\ref{MSCO-eq-MP2}).

\begin{figure}
\includegraphics[width=8cm]{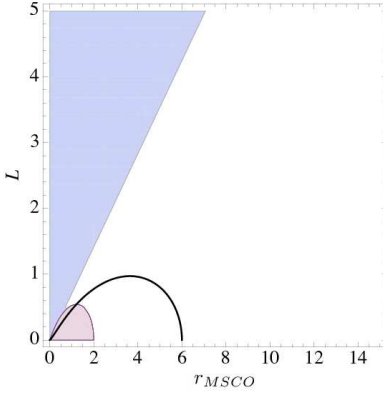}
\caption{ 
MSCOs for the Majumdar-Papapetrou solution, where 
we focus on two equal-mass black hole cases. 
The vertical axis denotes $L$ 
in the unit of $M$. 
The horizontal axis denotes 
the MSCO radius. 
The solid curve (black in color) is the MSCOs. 
The orbit is radially unstable in the inner area 
surrounded by the solid curve, 
while it is radially stable in the outer region. 
The lightly shaded region (blue in color) denotes the forbidden region 
where the orbit is unstable against the 
zenithal 
perturbation. 
In the dark shaded region (magenta in color), 
the orbit is unstable to the 
zenithal 
perturbation 
and $\varepsilon^2 < 0$ or $\ell^2 < 0$. 
Two MSCOs are possible, if $L$ is between $\sim 0.6$ 
and $\sim 1.0$. 
}
\label{fig-MP}
\end{figure}

The gravitational pull by the two black holes along the $z$ axis 
is unlikely to stabilize the particle position. 
Some of the equatorial circular orbits are indeed unstable 
against nonradial perturbations. 
See Figure \ref{fig-MP}.

\begin{figure}
\includegraphics[width=8cm]{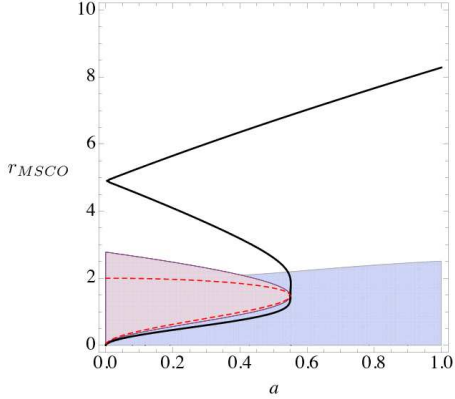}
\caption{ 
MSCOs in Johannsen and Psaltis's modified model of a rapidly spinning 
black hole. 
The vertical axis denotes the MSCO radius in the unit of $M$. 
The horizontal axis denotes 
the black hole spin parameter. 
The solid curve (black in color) denotes the ISCOs. 
The upper part of the solid curve ($r$ is larger than $\sim 5.6$) 
corresponds to the retrograde case 
and the lower part ($r$ is smaller than $\sim 5.6$) 
corresponds to the prograde one. 
The lightly shaded region (blue in color) 
denotes the forbidden region where 
the orbit is unstable against the 
zenithal 
perturbation. 
In the dark shaded region (magenta in color), 
the orbit is unstable against the 
zenithal perturbation 
and $\varepsilon^2 < 0$ or $\ell^2 < 0$. 
The other solid curve (red in color) denotes the horizon. 
There exist outer and inner horizons, 
when $a$ is smaller than $\sim 0.7 m$. 
They merge at $a \sim 0.7 m$. 
}
\label{fig-JP}
\end{figure}

\subsection{Johannsen and Psaltis's model for a modified Kerr metric}
Finally, we examine Johannsen and Psaltis's modified Kerr metric as \cite{JP} 
\begin{align}
ds^2=&-[1+h(r,\theta)]\left(1-\frac{2Mr}{\Sigma}\right)dt^2 
\notag\\
&-\frac{4aMr\sin^2\theta}{\Sigma}\times[1+h(r,\theta)]dtd\phi 
\notag\\
&+\frac{\Sigma[1+h(r,\theta)]}{\Delta+a^2\sin^2\theta h(r,\theta)}dr^2
+\Sigma d\theta^2 
\notag\\
&+\left[\sin^2\theta\left(r^2+a^2+\frac{2a^2Mr\sin^2\theta}{\Sigma}
\right)\right.
\notag\\
&~~~~~
\left.+h(r,\theta)\frac{a^2(\Sigma+2Mr)\sin^4\theta}{\Sigma}\right]d\phi^2 , 
\label{ds-Kerr-JP}
\end{align}
where we define 
\begin{align}
h(r,\theta) \equiv \epsilon_3\frac{M^3r}{\Sigma^2} . 
\end{align}
The new parameter $\epsilon_3$ is not constrained by observations 
so far \cite{JP}.

For the model proposed by Johannsen and Psaltis, 
Eq. (\ref{MSCO-eq}) becomes explicitly 
\begin{align}
&4 r^{22}\Big[9 a^4 + (6 M-r)^2 r^{2}-2 a^2 r (14 M+3 r)\Big] 
\notag\\
&+4 M^{2} r^{19} \epsilon_3 
\notag\\
&~~\times\Big[3 a^4 (4 M-15 r) 
\notag\\
&~~~~~~
+(6 M-r)\left(36 M^2-4 M r-3 r^2\right)r^{2}
\notag\\
&~~~~~~
+2 a^2  r \left(2 M^2-69 M r+3 r^2\right)\Big] 
\notag\\
&+M^{4} r^{15} \epsilon_3^2 
\notag\\
&~~\times\Big[-360 a^6 M + r^{3} \left(36 M^2-4 M r-3 r^2\right)^2
\notag\\
&~~~~~~
+4 M (6 M-r) r^{3} \left(96 M^2-74 M r+15 r^2\right) 
\notag\\
&~~~~~~
+3 a^4  r \left(224 M^2-696 M r+75 r^2\right)
\notag\\
&~~~~~~
+2 a^2 r^{2} \left(880 M^3-1344 M^2 r+102 M r^2+45 r^3\right)\Big]  
\notag\\
&+2 M^{7} r^{12} \epsilon_3^3 
\notag\\
&~~\times\Big[-1116 a^6 M +48 a^4 r \left(35 M^2-84 M r+15 r^2\right)
\notag\\
&~~~~~~
+ r^{3} \left(36 M^2-4 M r-3 r^2\right) \left(96 M^2-74 M r+15 r^2\right) 
\notag\\
&~~~~~~
+ a^{2} r^{2}\left(5056 M^3-6384 M^2 r+1854 M r^2-81 r^3\right)\Big] 
\notag\\
&+ M^{10} r^{9} \epsilon_3^4 
\notag\\
&~~\times\Big[-5976 a^6 M + r^{3} \left(96 M^2-74 M r+15 r^2\right)^2
\notag\\
&~~~~~~
+24 a^4 r \left(398 M^2-723 M r+171 r^2\right) 
\notag\\
&~~~~~~
+8 a^2 r^{2}\left(2432 M^3-3174 M^2 r+1227 M r^2-135 r^3\right)\Big] 
\notag\\
&+8 a^{2} M^{13} r^{6} \epsilon_3^5 
\notag\\
&~~\times\Big[-1017 a^4 M 
\notag\\
&~~~~~~
+6 a^2 r \left(284 M^2-435 M r+120 r^2\right)
\notag\\
&~~~~~~
+ r^2 \left(2344 M^3-3306 M^2 r+1515 M r^2-225 r^3\right)\Big] 
\notag\\
&+ 24 a^{4} M^{16} r^{3} \epsilon_3^6 
\notag\\
&~~\times\Big[-228 a^2 M 
+ r \left(404 M^2-525 M r+150 r^2\right)\Big] 
\notag\\
&- 1440 a^6 M^{20} \epsilon_3^7=0 . 
\end{align}

See Figure \ref{fig-JP} for a numerical result. 
Here, we choose $\epsilon_3 = 2$, 
such that the effects by the modification in the figure can be seen by eye.  
Circular orbits that 
are stable to radial perturbations for some parameter region 
are unstable to nonradial perturbations. 
This thing has not been known so far, 
because 
Johannsen and Psaltis \cite{JP} 
consider only the radial perturbations.

\section{Conclusion}
We performed 
a linear stability analysis to nonradial perturbations 
of a marginal stable circular orbit of a test particle 
in stationary axisymmetric spacetimes.
A zenithal stability criterion 
was obtained as 
Eq. (\ref{theta}). 
The proposed approach was applied to Kerr solution 
and Majumdar-Papapetrou solution to Einstein equation. 
Moreover, 
we suggested that, 
for 
the Johannsen and Psaltis's model, 
circular orbits that are stable to radial perturbations 
for some parameter region 
become unstable against 
zenithal 
perturbations. 
This implies that, 
at least for the Johannsen and Psaltis's model, 
the last circular orbit might be larger than the ISCO, 
because the ISCO (in the usual sense) often refers to 
the smallest circular orbit that  
is stable against the radial perturbation. 
\cite{Blanchet,Isoyama,Ono,Stute,BJS,RZ,Beheshti,Text}
It is left as a future work to study a more general case.

\begin{acknowledgments}
We are grateful to Luc Blanchet for his useful comments 
on the 3PN (and higher PN) ISCO. 
We wish to thank Takashi Nakamura, Kei Yamada, 
Yuuiti Sendouda and Yasusada Nambu for useful discussions. 
We would like to thank Takahiro Tanaka for his hospitality 
at the JGRG25 workshop in Kyoto, 
where this work was largely developed. 
This work was supported in part 
by JSPS Grant-in-Aid for Scientific Research, 
Kiban C, No. 26400262 (H.A.) and Shingakujutsu, No. 15H00772 (H.A.).
\end{acknowledgments}

\appendix
\section{Derivations of Eq. (\ref{MSCO-eq})} 
In this section, we discuss the derivation of Eq. (\ref{MSCO-eq}), 
where $\varepsilon\neq 0$ and $\ell \neq 0$.  
A method of deriving Eq. (\ref{MSCO-eq}) 
is using not only Eqs. (\ref{r-1}) and (\ref{r-2}) 
but also the equation defining the proper time along the particle orbit 
that is expressed as $-1 = A \dot{t}^2 + \cdots $. 
By using Eqs. (\ref{e}) and (\ref{l}), 
one can show that the three equations are 
linear in $\varepsilon^2$, $\ell^2$ and $\varepsilon\ell$. 
They are easily solved for $\varepsilon^2$, $\ell^2$ and $\varepsilon\ell$. 
A consistency condition as $(\varepsilon\ell)^2 = \varepsilon^2 \times \ell^2$ 
gives Eq. (\ref{MSCO-eq}). 

It is interesting to mention another but shorter derivation. 
Let $p$ and $q$ denote 
the left hand sides of Eqs. (\ref{r-1}) and (\ref{r-2}), respectively. 
They are polynomials in $\varepsilon$ and $\ell$ 
and 
therefore 
they are rearranged as 
\begin{align}
p \frac{\partial^2\tilde{D}}{\partial r^2} 
=& 
q \frac{\partial\tilde{D}}{\partial r} 
+ \left(
\frac{\partial\tilde{A}}{\partial r}
\frac{\partial^2\tilde{D}}{\partial r^2}
- \frac{\partial\tilde{D}}{\partial r}
\frac{\partial^2\tilde{A}}{\partial r^2} 
\right) 
\varepsilon^2 
\nonumber\\
&+ 
2 \left(
\frac{\partial\tilde{H}}{\partial r}
\frac{\partial^2\tilde{D}}{\partial r^2}
- \frac{\partial\tilde{D}}{\partial r}
\frac{\partial^2\tilde{H}}{\partial r^2}
\right)
\varepsilon\ell , 
\label{q}
\\
p \frac{\partial^2\tilde{A}}{\partial r^2} 
=& q \frac{\partial\tilde{A}}{\partial r}
+ 2 \left(
\frac{\partial\tilde{H}}{\partial r}
\frac{\partial^2\tilde{A}}{\partial r^2}
- \frac{\partial\tilde{A}}{\partial r}
\frac{\partial^2\tilde{H}}{\partial r^2}
\right) 
\varepsilon\ell
\nonumber\\
&+ 
\left(
\frac{\partial\tilde{A}}{\partial r}
\frac{\partial^2\tilde{D}}{\partial r^2}
- \frac{\partial\tilde{D}}{\partial r}
\frac{\partial^2\tilde{A}}{\partial r^2} 
\right) 
\ell^2 , 
\label{p}
\end{align}
Therefore, Eqs. (\ref{r-1}) and (\ref{r-2}) 
are equivalent to 
\begin{align}
&\left(
    \begin{array}{cc}
 \frac{\partial\tilde{A}}{\partial r}
\frac{\partial^2\tilde{D}}{\partial r^2}
- \frac{\partial\tilde{D}}{\partial r}
\frac{\partial^2\tilde{A}}{\partial r^2} 
&
2\frac{\partial\tilde{H}}{\partial r}
\frac{\partial^2\tilde{D}}{\partial r^2}
- 2\frac{\partial\tilde{D}}{\partial r}
\frac{\partial^2\tilde{H}}{\partial r^2}\\
2\frac{\partial\tilde{H}}{\partial r}
\frac{\partial^2\tilde{A}}{\partial r^2}
- 2\frac{\partial\tilde{A}}{\partial r}
\frac{\partial^2\tilde{H}}{\partial r^2} 
& 
 \frac{\partial\tilde{A}}{\partial r}
\frac{\partial^2\tilde{D}}{\partial r^2}
- \frac{\partial\tilde{D}}{\partial r}
\frac{\partial^2\tilde{A}}{\partial r^2}
   \end{array}
  \right)
\left(
    \begin{array}{c}
\varepsilon\\
l
  \end{array}
  \right)
\nonumber\\
&
=
\left(
    \begin{array}{c}
0\\
0
  \end{array}
  \right) .
\label{matrix-form}
\end{align}
A necessary condition for avoiding simultaneously $\varepsilon=0$ and $\ell=0$ 
is that the determinant of the matrix in the left hand side of 
Eq. (\ref{matrix-form}) vanishes. 
Eq. (\ref{MSCO-eq}) is thus derived.


\begin{thebibliography}{99}
\bibitem{Metric}
In this paper, we use the polar coordinates. 
In the cylindrical coordinates, the line element is known 
as the Weyl-Lewis-Papapetrou form \cite{Lewis,LR,Papapetrou}.
\bibitem{Reflection}
Note that this local condition suffices 
for the discussion in the present paper, 
though it does not guarantee the global existence of the spacetime.
\end{thebibliography}
\end{document}